\documentclass[a4paper,11pt]{article}

\usepackage[dvipsnames]{xcolor}
\usepackage{jheppub}
\usepackage{graphicx}
\usepackage{amsmath,amssymb,amsthm,amsfonts}
\usepackage{slashed}
\usepackage{braket}
\usepackage[export]{adjustbox}
\usepackage{enumitem}
\usepackage{placeins}
\usepackage[normalem]{ulem}
\usepackage{multirow}
\usepackage{hyperref}
\usepackage{mathrsfs}  
\usepackage{dsfont}
\usepackage{xcolor}
\usepackage{verbatim}
\usepackage[T1]{fontenc}
\usepackage[utf8]{inputenc}
\usepackage{array}
\usepackage{wrapfig}
\usepackage{tabu}
\usepackage{longtable}
\usepackage{float}
\usepackage{caption}
\usepackage{subcaption}
\usepackage{tikz,tikz-3dplot}
\usepackage[compat=1.0.0]{tikz-feynman}
\usetikzlibrary{calc,arrows.meta,shapes.geometric}
\usepackage{soul}
\usepackage[usestackEOL]{stackengine}
\usepackage{cancel}

\allowdisplaybreaks

\definecolor{mycolor}{rgb}{0.12, 0.56,0.33 }
\hypersetup{
    colorlinks=true,
    linkcolor=mycolor,
    citecolor=mycolor,
    filecolor=mycolor,      
    urlcolor=mycolor
    }

\def\beq{\begin{equation}}
\def\eeq{\end{equation}}
\def\bea{\begin{eqnarray}}
\def\eea{\end{eqnarray}}

\newcommand{\exbox}[1]{\begin{center}\fbox{\parbox[c]{14.8cm}{#1}}\end{center}}
\newcommand{\inline}[2]{\begin{tabular}{rl}{\tt \phantom{a}In[#1]:= } #2\end{tabular}}
\newcommand{\outline}[2]{\begin{tabular}{rl}{\tt Out[#1]:= } #2\end{tabular}}
\newcommand{\breakline}{\rule{14.8cm}{0.5pt}}

\definecolor{lightgrey}{RGB}{235,245,255}
\def\mybtab#1\myetab{\begin{tabular}{p{61mm}p{79mm}}#1\end{tabular}}
\def\btab#1\etab{\begin{tabular}{p{55mm}p{85mm}}#1\end{tabular}}
\def\btabx#1\etabx{\begin{tabular}{p{60mm}p{50mm}}#1\end{tabular}}
\def\btaby#1\etaby{\begin{tabular}{p{15mm}p{95mm}}#1\end{tabular}}
\def\bcen{\begin{center}\begin{small}}
\def\ecen{\end{small}\end{center}}
\def\mybgfb#1\myegfb{\bcen\fcolorbox{black}{lightgrey}{\parbox{148mm}{\mybtab#1\myetab}}\ecen}
\def\bgfb#1\egfb{\bcen\fcolorbox{black}{lightgrey}{\parbox{148mm}{\btab#1\etab}}\ecen}
\def\bgfbx#1\egfbx{\bcen\fcolorbox{black}{lightgrey}{\parbox{118mm}{\btabx#1\etabx}}\ecen}
\def\bgfbalign#1\egfbalign{\bcen\fcolorbox{black}{lightgrey}{\parbox{118mm}{\btaby#1\etaby}}\ecen}
\preprint{SLAC-PUB-250410}

\title{RVV$\times$V: Interference contributions to inclusive Higgs boson and Drell-Yan production at N$^4$LO in QCD}

\author[a]{Bernhard Mistlberger,}
\emailAdd{bernhard.mistlberger@gmail.com}
\author[a]{Adi Suresh}
\emailAdd{adisur@stanford.edu}

\affiliation[a]{SLAC National Accelerator Laboratory, Stanford University, Stanford, CA 94039, USA}

\abstract{
We present partonic contributions to the inclusive gluon-fusion Higgs boson and Drell-Yan production cross sections at Hadron colliders at next-to-next-to-next-to-next-to leading order (N$^4$LO).
Specifically, we compute contributions due to the interference of one-loop amplitudes with two-loop amplitudes with three QCD partons and the Higgs boson or a virtual photon. 
Our result is in the form of a Laurent expansion in the dimensional regulator $\epsilon$, the coefficients of which are analytic functions in the ratio of the Higgs boson or virtual photon mass to the partonic center of mass energy. 
Furthermore, we introduce and deploy a new package implemented in \textsc{Mathematica}, \textsc{Cifar} ("Color Invariant Feynman Amplitude Reducer"), to express color factors in terms of general Casimir invariants of simple compact Lie algebras. 
}

\begin{document}
\maketitle

\section{Introduction}

The discovery of the Higgs boson~\cite{Aad:2012tfa,Chatrchyan:2012xdj} at the Large Hadron Collider (LHC) and the absence the discovery of new elementary particles have paved the way for an era focused on precision collider physics. 
The LHC provides us with a remarkable window to uncover the profound mysteries of our current understanding of high energy physics and stringently test our theories of nature.
Precision Higgs-boson physics is at the very center of this exploration.
With the era of the high-luminosity LHC~\cite{ZurbanoFernandez:2020cco} quickly approaching and due to ever-improving experimental advancements, precision physics at the one-percent level will become a reality in the next decade for many observables. 
Deriving predictions based on our Standard Model (SM) of particle physics that rival the envisioned experimental challenge is a monumental task for the field of high energy theory.
In this article, we take a decisive first step to reduce theoretical uncertainties for some of the cornerstone observables of Higgs boson and electroweak precision physics.
Specifically, we begin the computation of next-to-next-to-next-to-next-to leading order (N$^4$LO) corrections in perturbative Quantum Chromodynamics (QCD) of the inclusive production rates of a Higgs boson via the gluon fusion mechanism and of a Drell-Yan lepton pair.

About 90\% of all Higgs bosons at the LHC are produced via the so-called gluon-fusion production mechanism~\cite{Georgi:1977gs}.
Consequently, very high precision predictions for this production mode are required to exploit LHC phenomenology at its fullest.
The strong coupling constant of QCD is the largest of all small coupling constants at the LHC and consequently significant effort was invested in the past to describe QCD corrections to this observable.
Corrections at next-to-leading order (NLO) were computed in refs.~\cite{Dawson:1990zj,Graudenz:1992pv,Spira:1995rr}. 
Next-to-next-to-leading order (NNLO) corrections were first calculated in refs.~\cite{Anastasiou2002,Harlander:2002wh,Ravindran:2003um} using an effective QCD theory, treating the top quark as infinitely massive and the other quarks as massless~\cite{Inami1983,Shifman1978,Spiridonov:1988md,Wilczek1977}. 
This approximation proves to be highly efficient, with corrections and even exact computations in full QCD becoming available~\cite{Czakon:2021yub,Harlander:2009mq,Pak:2009dg,Harlander:2009my,Marzani:2008az,Niggetiedt:2024nmp,Czakon:2024ywb,Harlander:2009bw,Czakon:2023kqm}.
The gluon-fusion cross section was the first LHC production cross section to be computed at next-to-NNLO (N$^3$LO), first as an expansion near the Higgs boson production threshold in ref.~\cite{Anastasiou:2015ema} and then exactly in ref.~\cite{Mistlberger:2018etf}.
Together with the computation of electroweak corrections in refs.~\cite{Actis:2008ts,Actis:2008ug,Aglietti:2004nj,Anastasiou:2008tj,Becchetti:2020wof,Bonciani:2022jmb,Bonetti:2017ovy,Bonetti:2018ukf}, these calculations enable today's precision comparisons of ATLAS and CMS data with SM predictions~\cite{Anastasiou:2016cez,LHCHiggsCrossSectionWorkingGroup:2016ypw}.
The residual uncertainty purely due to the truncation of the QCD perturbation series is at the level of $~2.5$\%~\cite{Dulat:2018rbf,Anastasiou:2016cez}, and thus on its own above or at the target precision of the LHC. 
Aside from uncertainties of parton distribution functions (which are of comparable size), it is the largest residual theoretical uncertainty.
This highly motivates us to go beyond the current paradigm and pursue a computation of the gluon fusion cross section at N$^4$LO.

The production cross section of a lepton pair in proton collisions - the so-called Drell-Yan (DY) process - is {\it the} standard candle process at the LHC. 
Measurable to astounding precision - it serves as a tool for calibration of the experiment, a means to extract Standard Model parameters like the weak mixing angle or the parton distribution functions of the proton, and as a test for physics beyond the Standard Model.
Achieving high-precision predictions for the DY process equally requires QCD perturbative corrections at very high order. 
The computation of the DY process can be based on the same tools and results as the computation of the gluon-fusion Higgs cross section and consequently, analytic results for this process are available today~\cite{Baglio:2022wzu,Duhr:2021vwj,Duhr:2020sdp,Duhr:2020seh}.
Furthermore, building on the same technology, it was possible to compute the production cross section of a Higgs boson in bottom quark fusion~\cite{Duhr:2019kwi,Duhr:2020kzd}, the cross section for the production of a vector boson which then radiates a Higgs boson~\cite{Baglio:2022wzu} and contributions to the di-Higgs-boson production cross section~\cite{Chen:2019fhs} at N$^3$LO.

Computing an LHC cross section to N$^4$LO is a daunting undertaking. 
In the case of the gluon-fusion cross section calculation at N$^3$LO, it proved a successful strategy to separate the computation of the partonic cross section into individual computations of different final-state parton multiplicity.
While these individual building blocks are not physical on their own, they separately represent gauge invariant and a well-defined approach that allows for the development of technology to complete the overarching goal successively. 
Purely virtual three-loop contributions were first computed in refs.~\cite{Gehrmann:2010tu,Gehrmann:2010ue,Baikov:2009bg,Gehrmann2005}.
Contributions due to one final-state parton and the interference of one-loop scattering amplitudes were computed in refs.~\cite{Kilgore:2013gba,Anastasiou:2013mca} and the interference of two-loop and tree-level amplitudes in refs.~\cite{Dulat:2014mda,Duhr:2014nda}. Contributions with two final-state partons were first computed in an expansion around the production threshold of the Higgs boson in refs.~\cite{Anastasiou:2015yha,Anastasiou:2015ema,Zhu:2014fma}, and similarly for three partons in the final state in refs.~\cite{Anastasiou:2014lda,Anastasiou:2015ema,Anastasiou:2013srw,Li:2014afw}. 
Ultimately, all contributions were computed exactly and combined in ref.~\cite{Mistlberger:2018etf}.
Today, purely virtual four-loop matrix elements are already available thanks to the computation of refs.~\cite{Lee:2022nhh,Agarwal:2021zft,Lee:2021uqq,vonManteuffel:2020vjv}.
Our aim is to continue following this successful strategy moving forward.

In this article, we present the calculation of a contribution to the N$^4$LO corrections to Drell-Yan and gluon-fusion Higgs-boson production. 
Specifically, we focus on the partonic cross section involving a single parton in the final state in addition to the color-singlet boson.
In addition, we restrict ourselves to the case of the interference of two-loop with one-loop scattering-matrix elements. 
We refer to this contribution as the "single-real double virtual cross single virtual" or $\text{RVV} \times \text{V}$ contribution.
While the required one- and two-loop amplitudes have been computed in the past~\cite{Badger:2006us,Badger:2009hw,Badger:2009vh,Dixon:2009uk,Gehrmann:2011aa,Gehrmann:2023etk}, we require these quantities in conventional dimensional regularization through very high order in the dimensional regulator. 
We perform the necessary computation of these scattering amplitudes and contract them with the required complex conjugate amplitudes to obtain our desired phase-space integrand. 
Carrying out the remaining phase-space integral over the degrees of freedom of the extra parton analytically represents a challenging task and we make use of cutting edge integration technology~\cite{Goncharov:2001iea,Panzer:2014caa,Panzer:2015ida,Maitre:2005uu,Duhr:2019tlz,Duhr:2011zq,Duhr:2012fh} to succeed in this step.
Our results are analytic functions given in terms of a Laurent series in the dimensional regulator expressed in terms of Goncharov polylogarithms~\cite{Goncharov:2001iea}. 
We make our results available in electronically readable form together with the arXiv submission of this article.

One ubiquitous step in the computation of QCD amplitudes and cross sections is the calculation of color factors. 
Typically, this involves the summation of generators of the different representations of the color group $SU(3)$ for individual Feynman diagrams.
However, it can be invaluable not only to work with an unspecified number of $N_c$ colors (instead of the physical value $N_c=3$), but also to generalize this step to any simple compact Lie group.
To streamline this step for high order computations we develop the package \textsc{Cifar} (Color Invariant Feynman Amplitude Reducer, pronounced "cipher"), implemented in \textsc{Mathematica}.
The implemented algorithm is based on ref.~\cite{vanRitbergen:1998pn}. 
Specifically. our package allows for the computation of fully contracted products of generators of a Lie group in terms of Casimir invariants. 
Our package was already applied to the computations in refs.~\cite{Herzog:2023sgb,Guan:2024hlf}.
We detail our algorithm and its implementation below and attach the package together with the arXiv submission of this article.

This paper is structured as follows. The general setup is introduced in section~\ref{sec:setup}. The details of calculating amplitudes and solving master integrals are outlined in section~\ref{sec:amplitudes}. Phase space integration and regulation of infrared singularities are described in section~\ref{sec:integration}. 
In section~\ref{sec:color} we describe our algorithm to compute color factors and its implementation in the package \textsc{Cifar}.
Finally, in section~\ref{sec:conclusions}, we conclude our discussion.
\newpage

\section{Calculation of $\text{RVV} \times \text{V}$ contributions to N$^4$LO Drell-Yan and Higgs-boson production} 

One of the main results of this article is the computation of contributions to the inclusive, partonic production cross sections of a Higgs boson or Drell-Yan pair due to the interference of two-loop with one-loop scattering amplitudes. 
We denote this contribution to the partonic cross section as "single-real double virtual cross single virtual" or $\text{RVV} \times \text{V}$.
We start this section by providing a general setup and definition for these contributions.
Next, we discuss the analytic computation of the scattering amplitudes that are the ingredients for our result. 
Finally, we discuss the interference and integration of these scattering amplitude over the one-parton phase space. 
  
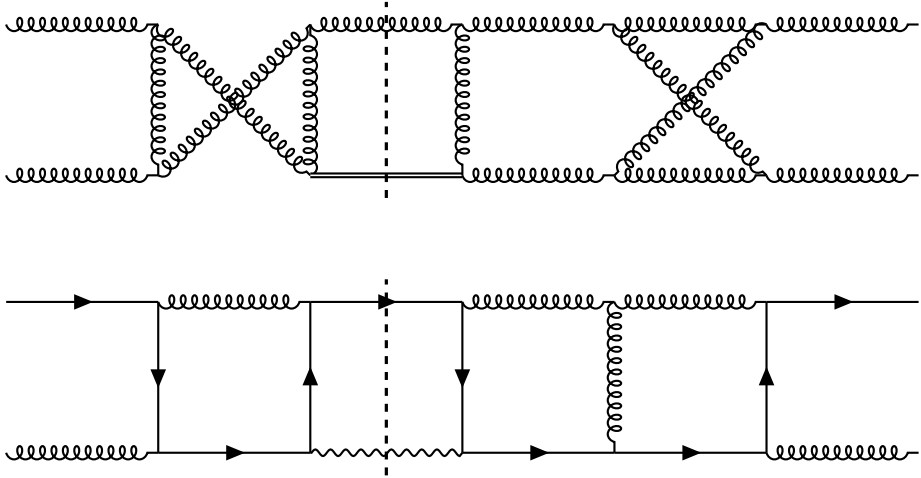
\begin{figure}[H]
    \centering
    \begin{tikzpicture}
        \begin{feynman}[large]
        
        \vertex (1);
        \vertex[below=of 1] (2);
        
        \vertex[right=of 1] (156);
        \vertex[right=of 2] (257);
        \vertex[right=of 156] (378);
        \vertex[right=of 257] (468);

        \vertex[right=of 378] (3911c);
        \vertex[right=of 468] (41011c);

        \vertex[right=of 3911c] (689c);
        \vertex[right=of 41011c] (7810c);
        \vertex[right=of 689c] (156c);
        \vertex[right=of 7810c] (257c);

        \vertex[right=of 156c] (1c);
        \vertex[right=of 257c] (2c);

        \diagram*{
            (1) --[gluon] (156),
            (2) --[gluon] (257),
            (156) --[gluon] (257),
            (156) --[gluon] (468),
            (257) --[gluon] (378),
            (468) --[gluon] (378),

            (378) --[gluon] (3911c),
            (468) --[double] (41011c),
            
            (3911c) --[gluon] (41011c),
            (3911c) --[gluon] (689c),
            (41011c) --[gluon] (7810c),
            (7810c) --[gluon] (257c),
            (689c) --[gluon] (156c),
            (156c) --[gluon] (7810c),
            (689c) --[gluon] (257c),
            (156c) --[gluon] (1c),
            (257c) --[gluon] (2c),
            
            };
        \end{feynman}

        \draw[dashed, very thick] (5, -2.3) -- (5, 0.3);
        
    \end{tikzpicture}

    \vskip 1cm
    
    \begin{tikzpicture}
        \begin{feynman}[large]
        
        \vertex (1);
        \vertex[below=of 1] (2);
        
        \vertex[right=of 1] (156);
        \vertex[right=of 2] (257);
        \vertex[right=of 156] (368);
        \vertex[right=of 257] (478);

        \vertex[right=of 368] (3911c);
        \vertex[right=of 478] (41011c);

        \vertex[right=of 3911c] (689c);
        \vertex[right=of 41011c] (7810c);
        \vertex[right=of 689c] (156c);
        \vertex[right=of 7810c] (257c);

        \vertex[right=of 156c] (1c);
        \vertex[right=of 257c] (2c);

        \diagram*{
            (1) --[fermion] (156),
            (2) --[gluon] (257),
            (156) --[fermion] (257),
            (156) --[gluon] (368),
            (257) --[fermion] (478),
            (468) --[fermion] (368),

            (368) --[fermion] (3911c),
            (468) --[photon] (41011c),
            
            (3911c) --[fermion] (41011c),
            (3911c) --[gluon] (689c),
            (41011c) --[fermion] (7810c),
            (689c) --[gluon] (7810c),
            (7810c) --[fermion] (257c),
            (689c) --[gluon] (156c),
            (257c) --[fermion] (156c),
            (156c) --[fermion] (1c),
            (257c) --[gluon] (2c),
            
            };
        \end{feynman}

        \draw[dashed, very thick] (5, -2.3) -- (5, 0.3);
        
    \end{tikzpicture}

    \caption{An example contribution to the Higgs (Drell-Yan) cross section via $g+g \rightarrow H+g$ ($q g \rightarrow \gamma^* + q$) is shown on the top (bottom). The dashed vertical lines represent phase space integration, and the double line represents the Higgs boson. The propagators cut by the dashed line are on-shell (or set to a fixed off-shell mass $Q$ in the case of the photon.) The left side of the dashed lines show contributions to the one-loop amplitude, and the right side of the dashed lines show contributions to the (complex-conjugated) two-loop amplitude.}
    \label{fig:RVVxVExampleDiagram}
\end{figure}
    
\subsection{Setup}
\label{sec:setup}

The focus of this article are QCD corrections to the hadron collider processes for inclusive production of a color singlet $B$ (i.e. the Higgs boson or a virtual photon) with invariant mass $Q$.

\beq
\text{Proton}(P_1)+\text{Proton}(P_2) \rightarrow B(Q) + X,
\eeq
where the incoming protons have momenta $P_1$ and $P_2$. Using the parton model and factorization of long and short interactions via the partonic cross section $\hat \sigma_{i j \rightarrow B+X}$ and parton distribution functions $f_i$, we can express the hadronic cross section as a sum over all contributing partonic processes: 
\beq
\sigma_{PP\rightarrow B+X} = \tau \sum_{i,j} \int_\tau^1 \frac{dz}{z} \int_\frac{\tau}{z}^1 \frac{d x_1}{x_1} f_i(x_1) f_j \left(\frac{\tau}{x_1 z}\right)\frac{1}{z}\hat \sigma_{ij \rightarrow B+X}(z,Q^2).
\eeq
The arguments of the parton distributions functions are the momenta fractions of the parton momenta to the proton momenta, $p_i=x_i P_i$. Here, we defined the center-of-mass energies and various ratios:
\bea
S &= (P_1+P_2)^2, \hspace{1cm}\tau &= \frac{Q^2}{S}, \\
s &= (p_1+p_2)^2,\hspace{1cm}z&= \frac{Q^2}{s},
\eea
which relate $x_2 = \frac{\tau}{x_1 z}$. 

In the case of Higgs production, we employ an effective field theory in which the top quark has infinite mass and has been integrated out, and all other quarks are massless~\cite{Inami1983,Shifman1978,Spiridonov:1988md,Wilczek1977}. 
Then, the Higgs appears to directly couple to gluons with a dimension-five effective operator~\cite{Kataev:1981gr,Chetyrkin:1997un,Kramer:1996iq,Schroder:2005hy,Chetyrkin:2005ia}.
The associated Lagrangian is:
\beq
\mathcal{L}_{\text{EFT}}=\mathcal{L}_{\text{SM},5}-\frac{1}{4}C^0 H G^a_{\mu \nu} G_a^{\mu \nu},
\eeq
where $\mathcal{L}_{\text{SM},5}$ is the Standard Model Lagrangian with five massless quarks and $C^0$ is the Wilson coefficient serving as the effective coupling between the Higgs and gluons.

Then, we define the so-called partonic coefficient functions $\eta^{(n)}_{i j}$ and relate them to the partonic cross section:
\beq
\label{eq:PCFdef}
\frac{1}{z}\hat \sigma_{ij \rightarrow B+X}(z,Q^2) = \hat{\sigma}_{B,0} \sum^\infty_{n=0} \left(\frac{\alpha_S^0}{\pi}\right)^n \eta^{(n)}_{i j}(z),
\eeq
where $\hat{\sigma}_{B,0}$ are the Born cross sections:
\beq
\hat{\sigma}_{\text{Higgs},0}= \frac{\pi C_0^2}{8 D_A}, \quad  \hat{\sigma}_{\text{DY},0}= \frac{\pi (q_i e)^2}{Q^2 D_F}.
\eeq
Above,  $D_A$ and $D_F$ are the dimensions of the QCD gauge group in the adjoint and fundamental representation respectively and are related to the number of colors $n_c$ as follows.
\beq
\label{eq:SUncDims}
D_A=n_c^2-1,\hspace{1cm}D_F=n_c.
\eeq
Above, $e$ is the electromagnetic coupling and $q_i$ is the fractional electric charge of parton $i$.
We determine the bare partonic coefficient functions order-by-order by computing:
\beq
\Tilde{\eta}^{(n)}_{ij}(z) = \frac{N_{i j}}{2 Q^2 \hat{\sigma}_{B,0}} \sum_{m=0}^n \int d \Phi_{B + m} \mathcal{M}^{(n)}_{i j \rightarrow B + m}.
\eeq
The bare partonic coefficient function is related to the renormalized partonic coefficient function of eq.~\eqref{eq:PCFdef} via renormalization of ultraviolet singularities.
The initial state averaging factors over spin/polarization and color are given by:
\bea
N_{gg} = \frac{1}{4(1-\epsilon)^2 D_A^2}, \\
N_{gq}  =N_{qg} = \frac{1}{4(1-\epsilon) D_A D_F}, \\
N_{q \bar{q}} = N_{q q} = N_{q Q} = \frac{1}{D_F^2},
\eea
where $g$, $q$, $\bar{q}$, and $Q$ represent gluon, quark, anti-quark, or differently-flavored quark in the initial state. 
$d\Phi_{B+m}$ is the phase space measure for the color singlet $B$ and $m$ additional partons.
$\mathcal{M}^{(n)}_{i j \rightarrow B + m}$ represents the coefficient of $\alpha_S^n$ in the expansion of the coupling constant for the squared modulus of all amplitudes where partons $i$ and $j$ produce said color singlet along with $m$ partons, summed over all polarizations and colors. 
Throughout our computation we work in the framework of dimensional regularization by extending the spacetime dimension $d$ by a small regulator $\epsilon$,such that $d=4-2\epsilon$.

To calculate the partonic coefficient functions at $n^{\text{th}}$ order, we need every combination of $l$-loop matrix elements involving $m$ external particles such that $m + l = n$. 
In this work, we consider the contribution with one additional external parton ($m=1$).
The contribution of the squared scattering-matrix element at $n^{\text{th}}$ power in the coupling is given by the following interference of scattering amplitudes $\mathcal{A}_X$.
\beq
\mathcal{M}^{(n)}_{i j \rightarrow B + k}=\sum_{l=0}^n \mathcal{A}^{(l)}_{ij\to B+k}\mathcal{A}^{\dagger \, (n-l)}_{ij\to B+k}.
\eeq
Specifically, at $n=3$ we find 
\beq
\mathcal{M}^{(3)}_{i j \rightarrow B + k}=2 \Re{\left[A^{(0)}_{ij\to B+k}A^{\dagger\, (3)}_{ij\to B+k}\right]} +  2\Re{\left[A^{(1)}_{ij\to B+k} A^{\dagger\,(2)}_{ij\to B+k} \right]}. 
\eeq
The second term in the above equation contains one-loop amplitudes interfered with two-loop amplitudes. 
The partonic cross section resulting from these amplitudes is the main result of this article and we refer to these contributions as the "single-real double virtual cross single-virtual" or $\text{RVV} \times \text{V}$ contribution.
Examples of such contributions are presented in fig.~\ref{fig:RVVxVExampleDiagram}.
\beq
\Tilde{\eta}^{\text{RVV}\times \text{V}}_{ij}(z) = \frac{N_{i j} (4\pi)^{-4 \epsilon} e^{4 \epsilon \gamma_E} }{2 Q^2 \hat{\sigma}_{B,0}}  \int d \Phi_{B + 1}  2\Re{\left[A^{(1)}_{ij\to B+k} A^{\dagger\,(2)}_{ij\to B+k} \right] }.
\eeq
Note, that we included in the above equation the usual factors ($(4\pi)^{-4 \epsilon} e^{4 \epsilon \gamma_E}$) associated with the modified minimal-subtraction scheme ($\overline{\text{MS}}$ scheme), where $\gamma_E$ is the Euler-Mascheroni constant.
We compute all partonic channels required for a complete computation of this contribution to the N$^4$LO production cross section of a Higgs boson or a Drell-Yan pair.
\bea
g + g &\rightarrow H + g, \\
q + \bar{q} &\rightarrow H + g,\\
q + g &\rightarrow H + q, \\
g + g &\rightarrow \gamma^* + g, \\
q + \bar{q} &\rightarrow \gamma^* + g,\\
q + g &\rightarrow \gamma^* + q.
\eea
We proceed by first computing the relevant scattering amplitudes and then integrate the required interfered amplitudes over the one parton phase space.
We provide further details below.

\subsection{Amplitudes}
\label{sec:amplitudes}

In this section, we discuss the computation of various four-point tree-level, one- and two- loop amplitudes. 
In the literature, many of the relevant amplitudes were already computed at least to some order in the dimensional regulator.
One- and two-loop amplitudes were obtained in refs.~\cite{Gehrmann:2023etk,Gehrmann:2022vuk,Gehrmann:2023zpz,Gehrmann:2013vga,Gehrmann:2011aa,Garland:2001tf,Garland:2002ak}
The three-loop amplitude is currently known in the soft~\cite{Chen:2023hmk,Herzog:2023sgb} and collinear~\cite{Guan:2024hlf} limits.
The leading color contribution for the production of a virtual photon at three loops was presented in ref.~\cite{Gehrmann:2023jyv} and for the production of a Higgs in ref.~\cite{Chen:2025utl}.
For the purposes of this article are perform the computation of the required one-loop and two-loop amplitudes and obtain results through transcendental-weight eight. 

Throughout our computation, we work in conventional dimensional regularization. 
First, we identify a set of gauge-invariant Lorentz tensor structures to express our scattering amplitudes. 
Next, we create Lorentz projectors onto these tensor structures (see refs.~\cite{Gehrmann:2013vga,Gehrmann:2011aa,Peraro:2020sfm,Goode:2024cfy,Goode:2024mci} for examples and recent developments).
We generate all relevant Feynman diagrams for our scattering amplitudes with \textsc{Qgraf}~\cite{qgraf}. 
Then, we apply our Lorentz projectors to the Feynman diagrams using a private C++ code based on \textsc{GiNaC}~\cite{Bauer2000} and perform spinor and tensor algebra.
We project the color structure on a basis of color tensors using our new package \textsc{Cifar}, which we describe in detail below.
As a result, we obtain a scalar integrand for the coefficients of our gauge-invariant tensor structures.

We proceed using standard loop integration technology to then compute the amplitudes. 
We use integration-by-part (IBP) identities~\cite{Laporta:2001dd,Chetyrkin1981,Tkachov1981} using a private implementation of Laporta's algorithm to relate the scalar Feynman integrals of our amplitudes to a basis of so-called master integrals. 
Next, we compute the master integrals using the method of differential equations~\cite{Henn:2013pwa,Gehrmann:1999as,Kotikov:1990kg,Kotikov:1991hm,Kotikov:1991pm} and transform the master integrals to so-called canonical form~\cite{Henn:2013pwa} using algorithm techniques outlined in ref.~\cite{Lee:2014ioa}.
We constrain boundary conditions for the differential equations using regularity conditions and consistency relations (see for example refs.~\cite{Henn:2020lye,Dulat:2014mda,Henn:2013nsa}). 
We obtain a solution to our master integrals in terms of generalized polylogarithms~\cite{Goncharov1998} through transcendental-weight eight.
Throughout our computation, we rely on the advancements made by the field for the computation of these special functions - see for example refs.~\cite{Huber:2007dx,Huber:2005yg,Maitre:2005uu,Remiddi:1999ew,Duhr:2011zq,Duhr:2012fh,Duhr:2019tlz}.
We then insert our solution for the master integrals in our newly computed amplitudes.  

To validate this step of our computation, we check that the infrared and collinear poles of our amplitudes match the prediction as derived in refs.~\cite{Almelid:2015jia,Aybat:2006mz,Aybat:2006wq,Catani:1998bh,Dixon:2008gr,Korchemsky:1987wg,Sterman:2002qn,Becher:2019avh}.
Second, we check that the soft and collinear limits of our scattering amplitudes factorize into the universal building blocks as computed in refs.~\cite{
Weinberg:1965nx,
Bern:1995ix,
Bern:1998sc,
Bern:1999ry,
Catani:2000pi,
Badger:2004uk,
Li:2013lsa,
Duhr:2013msa,
Dixon:2019lnw,
Kosower:1999rx,
Bern:2004cz,
Guan:2024hlf,
Herzog:2023sgb}.

\subsection{Phase Space Integration}
\label{sec:integration}

We label the momenta of our four-point process for partons $i$ and $j$ to a massive color-singlet $B$ and an additional parton $X$ follows:
\beq
i(p_1) + j(p_2) \rightarrow X(p_3) + B(p_4).
\eeq
This defines the usual kinematic invariants:

\beq
s \equiv 2 p_1 \cdot p_2 + i0, \quad t \equiv 2 p_2\cdot p_3 - i0, \quad u \equiv 2 p_1\cdot p_3 - i0,
\eeq
where we included the infinitesimal imaginary parts inherited from the Feynman prescription.

Working in dimensional regularization with $d=4-2 \epsilon$ dimensions, the two-particle outgoing phase space is given by:
\beq
d \Phi_2 = (2 \pi)^d \delta^{(d)}(p_1 +p_2 - p_3 - p_4) \frac{d^dp_3}{(2\pi)^{d-1}} \delta_+\left(p_3^2\right) \frac{d^dp_4}{(2\pi)^{d-1}} \delta_+\left(p_4^2 -Q^2\right).
\eeq
We further parameterize our kinematics from $s$, $t$, and $u$ to new variables $\lambda$, $\bar{z}\equiv 1-z$, and $Q^2$:
\beq
s =\frac{Q^2}{1 - \bar{z}}, \quad t=Q^2\left(\frac{ \bar{z}}{1-\bar{z}}\right) \lambda,\quad u=Q^2\left(\frac{ \bar{z}}{1-\bar{z}}\right)\left(1- \lambda\right).
\eeq
Then, performing the integral over the phase space of $B$ using the momentum-conserving Dirac delta function and applying spherical symmetry simplifies the phase space measure to:
\beq
d \Phi_2 = \frac{(4 \pi)^\epsilon \left(Q^2\right)^{-\epsilon} \left(1-\bar{z}\right)^\epsilon \bar{z}^{1-2\epsilon}}{8 \pi \, \Gamma(1-\epsilon)} d\lambda \left[ \lambda (1-\lambda)\right]^{-\epsilon} \Theta(\lambda)\Theta(1-\lambda).
\eeq

Note, that the collinear limits of parton $X$ becoming (anti-)paralell to the beam line are characterized by $\lambda\rightarrow 0$ and $\lambda\rightarrow 1$. 
For partonic coefficient functions with collinear singularities, integration over $\lambda$ results in additional poles in the dimensional regulator $\epsilon$. The soft limit of $X$ is characterized by $\bar{z} \rightarrow 0$. We regulate partonic coefficients functions with soft singularities by expanding them in terms of Dirac delta functions and plus distributions:
\beq
\left(\frac{1}{\bar{z}}\right)^{1 + a \epsilon} =- \frac{1}{a \epsilon}\delta(\bar{z}) + \sum_{k=0}^\infty\frac{(- a \epsilon)^k}{k!} \left[\frac{\log^k \bar{z}}{\bar{z}}\right]_+,
\eeq
where the plus distribution is defined (for a test function $f(x)$):
\beq
\int_0^1 dx \left[\frac{\log^k x}{x}\right]_+ f(x) \equiv \int_0^1 \frac{\log^k x}{x} \left( f(x) - f(0)\right).
\eeq

After parametrization and regulation of soft and collinear singularities, we perform integration over $\lambda$ order-by-order in the dimensional regulator $\epsilon$ using a private code implemented in \textsc{Mathematica} that incorporates \textsc{PolyLogTools}~\cite{Duhr:2019tlz}. 
Our results are a Laurent expansion in $\epsilon$ up to finite order in terms of rational functions of $\bar{z}$, generalized polylogarithms in $\bar{z}$, and Dirac delta functions and plus distributions in $\bar{z}$. 

As a check, we compared the soft-singular part of our results for $g + g \rightarrow H+g$ and $q +\bar{q} \rightarrow \gamma^* + g$ with what one obtains by doing the calculation after applying the single-emission soft current~\cite{Li:2013lsa,Duhr:2013msa} to the relevent three-point amplitudes obtained in refs.~\cite{Lee:2022nhh,Agarwal:2021zft,Lee:2021uqq,vonManteuffel:2020vjv,Gehrmann:2010ue,Gehrmann:2010tu,Baikov:2009bg,Gehrmann2005} and found full agreement. 
Furthermore, we validated our approach by re-computing all lower order contributions to the inclusive Higgs boson and Drell-Yan production cross section involving a single final state parton~\cite{Duhr:2014nda,Kilgore:2013gba,Anastasiou:2013mca,Dulat:2014mda} and find full agreement.

The results of this paper are provided in electronically readable form as ancillary files to the arXiv submission of this article. 
Specifically, we provide all necessary $\text{RVV}\times \text{V}$ partonic coefficient functions required for the complete computation of the Higgs boson and Drell-Yan production cross section through N$^4$LO. 
Furthermore, we provide lower order one-parton final state contributions to the partonic coefficient function to higher order in the dimensional regulator.

\section{CIFAR: A Color Algebra Package}
\label{sec:color}
In this section, we introduce \textsc{Cifar} (Color Invariant Feynman Amplitude Reducer, pronounced "cipher"), a \textsc{Mathematica} package that computes the color factors of a Feynman diagram in terms of Casimir invariants of a simple compact Lie algebra. We also provide two worked examples and sample usage in \textsc{Mathematica}. 
\textsc{Cifar} will find its application in computations through N$^4$LO in QCD and can compute color factors that include quartic Casimir invariants.

\subsection{Background and Definitions}
We consider Lie algebras whose generators $t^a$ follow the canonical commutation relation according to structure constants $f^{a b c}$:

\beq
\label{eq:lieCommutation}
\left[t^a_R,t^b_R\right]=i f^{a b c} t^c_R,
\eeq
where $R$ is the representation index (e.g. $F$ for fundamental and $A$ for adjoint).
Note that, here and throughout this section, repeated indices are implicitly summed over. We define fully symmetric color tensors in a given representation with the symmetrized trace:
\bea
\label{eq:symcolortensors}
d_R^{a_1 \dots a_n} & \equiv&  \text{Str}\left[t_R^{a_1}\dots t_R^{a_n}\right] \\ &=&\frac{1}{n!} \sum_\pi \text{Tr} \left[t_R^{a_{\pi(1)}}\dots t_R^{a_{\pi(n)}}\right],
\eea
where the sum includes all permutations of the generator indices $\{a_i\}$ when taking the trace of the product of the generators. Then, general Casimir invariants are defined as contractions of these tensors:
\beq
\label{eq:generalCasimir}
C^{R_1R_2}_{n} = d_{R_1}^{a_1 \dots a_n}d_{R_2}^{a_1 \dots a_n}.
\eeq
In particular, we identify the so-called quadratic Casimir invariants as:
\beq
\label{eq:quadraticCasimir}
\left(t_R^a t_R^a \right)_{i j} = C_R \delta_{i j}.
\eeq

The structure constants themselves satisfy the Lie algebra; this is the "adjoint" representation:
\beq
(t^a_A)_{b c} \equiv -i f^{a b c},
\eeq
Note, since the structure constants are antisymmetric by definition, we can swap the generator index and the entry indices in this representation simply at the cost of a minus sign. The adjoint quadratic Casimir is given by:
\beq
\label{eq:adjointQuadraticCasimir}
f^{a c d}f^{b c d} = C_A \delta^{a b}.
\eeq
For the algebra $\text{SU}(n_c)$, these quantities are:
\beq
\label{eq:SUncCasimir}
\begin{aligned}
 &C_F = \frac{n_c^2-1}{2n_c},\hspace{1cm} C_A=n_c,\hspace{1cm} \\
 &C_3^{FF} =\frac{4-5 n_c^2+n_c^4}{16 n_c},\hspace{1cm } C_3^{AF}=0, \hspace{1cm} C_3^{AA}=0,\\
 &C_4^{FF}=\frac{-18+24 n_c^2-7 n_c^4+n_c^6}{96 n_c^2},~ C_4^{AF}=\frac{n_c\left(-6+5 n_c^2+n_c^4\right)}{48}  , ~ C_4^{AA}= \frac{n_c^2\left(-36+35 n_c^2+n_c^4\right)}{24}.
\end{aligned}
\eeq
Additionally, the dimensions of the fundamental and adjoint representation in $\text{SU}(n_c)$ are given in eq.~\ref{eq:SUncDims} and are repeated here:
\beq
\label{eq:SUncDims2}
D_F=n_c, \hspace{1cm} D_A=n_c^2-1.
\eeq
Moreover, in $\text{SU}(n_c)$, one can use the canonical commutation relation (eq.~\ref{eq:lieCommutation}) to express all structure constants in terms of traces of products fundamental generators:
\beq
\label{eq:ftot}
f^{a b c} = 2i\left(\text{Tr} [t_F^a t_F^c t_F^b ]-\text{Tr} [t_F^a t_F^b t_F^c]\right).
\eeq
Then, in combination with the Fierz Identity 
\beq
\label{eq:Fierz}
t^a_{i j} t^a_{k l} = \frac{1}{2}\left(\delta_{i l}\delta_{jk}-\frac{1}{n_c} \delta_{i j} \delta_{kl}\right),
\eeq
one can express all contracted expressions in the color group in terms of $n_c$.

Scattering amplitudes of QCD partons are tensors in color space. 
To compute scalar amplitudes, it is often convenient to project on a basis of color tensors. 
The projection is achieved by contracting with a basis of color tensors structures yielding scalar color factors.
We can express all color factors in terms of general Casimir invariants and dimensions of various representations of the color gauge group. 
In the following, we outline our algorithm to achieve this goal.

For products of generators in the fundamental representation, we introduce the following shorthand notation.
\beq
\label{eq:fundProduct}
T^{a_1 \dots a_n}_{i j} \equiv \left(t_F^{a_1} \dots t_F^{a_n}\right)_{i j}.
\eeq
Furthermore, when these products are traced over, we omit the entry indices, 
\begin{align}
\label{eq:fundLoop}
T^{a_1 \dots a_n} &\equiv T^{a_1 \dots a_n}_{i i} \\ &=\text{Tr}\left[t_F^{a_1} \dots t_F^{a_n}\right].   
\end{align}

In case of the adjoint representation, structure constants themselves are commonly used to express color factors of amplitudes instead of the explicit adjoint generators. 
We will refer to the first index in a particular structure constant as the "loop" index, and the following two indices as "tracing" indices.
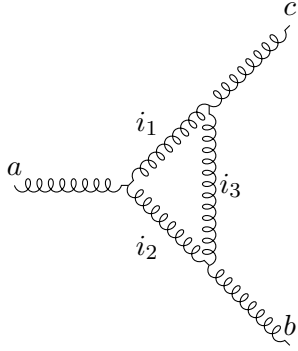
\begin{figure}[H]
    \centering

    \begin{tikzpicture}
    \begin{feynman}
        \vertex[label=\(a\)] (a);
        \vertex[right=of a] (a12);
        \vertex[above right=of a12] (c31);
        \vertex[below right=of a12] (b23);
        \vertex[below right=of b23, label=\(b\)] (b);
        \vertex[above right=of c31, label=\(c\)] (c);
        
        \diagram*[]{
            (a) --[gluon] (a12),
            (a12) --[gluon, edge label'=\(i_2\)] (b23),
            (b23) --[gluon, edge label'=\(i_3\)] (c31),
            (c31) --[gluon, edge label'=\(i_1\)] (a12),
            (c31) --[gluon] (c),
            (b23) --[gluon] (b),
            
        };
    \end{feynman}
\end{tikzpicture}
    \caption{Three gluon loop; the external gluons have adjoint indices $a$, $b$, and $c$, while the internal gluons are labeled $i_1$, $i_2$, and $i_3$.}
    \label{fig:gluon loop}  
\end{figure}
For example, consider fig.~\ref{fig:gluon loop} which has the color factor $f^{a i_1 i_2} f^{b i_2 i_3} f^{c i_3 i_1}$.
Here, $a$, $b$, and $c$ are loop indices, and $i_1$, $i_2$, and $i_3$ are tracing indices. 
Similar to eq.~\ref{eq:fundLoop}, we introduce the following notation for adjoint "loop" traces denoted by $F$.
The indices of $F$ display explicitly the loop indices of these adjoint traces. 
\beq \label{F}
F^{a b c} \equiv f^{a i_1 i_2} f^{b i_2 i_3} f^{c i_3 i_1},
\eeq
and more generally:
\beq \label{eq:Floop}
F^{a_1 a_2 \dots a_n} \equiv f^{a_1 i_1 i_2} f^{a_2 i_2 i_3} \dots f^{a_n i_n i_1}.
\eeq
When possible, we always commute indices such that loop indices are first and tracing indices follow in the same form as eq. \ref{eq:Floop}.

\subsection{Reduction for Fundamental Traces}
\label{sec:ReduceTT}
For traces in the fundamental representation, we first eliminate contracted adjoint indices in the same trace. 
We use the commutation relation eq.~\ref{eq:lieCommutation} to commute the indices until they are close (or if they are already neighboring) and then employ the following identities.
\bea
\label{eq:ContractTT1}
T^{\dots a b a \dots} &=& \left(C_F - \frac{1}{2}C_A\right) T^{\dots b \dots}, \\
\label{eq:ContractTT2}
T^{\dots b a a c \dots} &=& C_F T^{\dots b c \dots}.
\eea
Then, we further commute indices in the trace such that those that are doubly contracted with the same structure constant are neighboring; we then use:
\beq
\label{eq:ContractffTT}
f^{a b c}T^{\dots b c\dots} =\frac{i}{2} C_A T^{\dots a\dots}.
\eeq
Finally, after all indices within the trace are unique, we reduce it to color tensors $d$ and structure constants $f$, iterating the equation:
\beq
T^{a_1 \dots a_n}= T^{a_1 \dots a_n}- \frac{1}{n!} \sum_\pi T^{a_{\pi(1)} \dots a_{\pi(n)}} + d_F^{a_1 \dots a_n}. 
\eeq
We commute indices in each term within the symmetrized trace such that their sum cancels out the original trace. The commutator terms that are left over will be in terms of smaller fundamental traces and structure constants. The recursion terminates with 
\beq
\label{eq:quadraticDynkinIndex}
T^{a_1 a_2}=T_F\delta^{a_1 a_2},
\eeq
where $T_F$ is the quadratic Dynkin index of the fundamental representation, commonly chosen 1/2. 
Contracting this equation with an additional $\delta^{a_1 a_2}$ and applying eq.~\ref{eq:quadraticCasimir} yields the following identity:
\beq
C_F D_F=T_F D_A,
\eeq
which can be used to eliminate either the dimension of the fundamental or adjoint representation from all results if desired.
We assert the latter in the results presented in this paper.

To increase the speed of computing the terms in the symmetrized trace, we make use of the following simplifications.
\begin{enumerate}
\item 
Since a trace is invariant under cyclic permutations, we can factor out identical terms as:
\beq
\frac{1}{n!} \sum_\pi T^{a_{\pi(1)} \dots a_{\pi(n)}} = \frac{1}{(n-1)!} \sum_\pi T^{a_1 a_{\pi(2)} \dots a_{\pi(n)}},
\eeq
where now the sum is over all permutations of the latter $n-1$ indices; now we have a factor of $n$ fewer terms to commute, and the first index $a_1$ is always in the correct position.  
\item 
We tabulate the expressions for three- and four- index traces:
\beq
\label{eq:TrTabc}
T^{a_1 a_2 a_3}=d_F^{a_1 a_2 a_3}+ \frac{i}{2} T_F f^{a_1 a_2 a_3},
\eeq
\begin{equation}
\begin{split}
T^{a_1 a_2 a_3 a_4} =\,&d_F^{a_1 a_2 a_3 a_4}+ \frac{i}{2} \left(d_F^{a_1 a_4 b} f^{a_2 a_3 b} - d_F^{a_2 a_3 b} f^{a_1 a_4 b}\right)\\ &+\frac{1}{6}T_F \left(f^{a_1 a_4 b} f^{a_2 a_3 b}-f^{a_1 a_2 b} f^{a_3 a_4 b}\right).  
\end{split}
\end{equation}
\end{enumerate}
\subsection{Adjoint Traces}
\label{sec:AdjointReduction}
\subsubsection{Algorithm for Identifying Traces in the Adjoint Representation}
\label{sec:IDAdjointTrace}
Traces of structure constants are harder to identify as they carry three adjoint indices.
Since we consider color factors with fully contracted indices only and indices in the adjoint representation can be swapped (at the cost of a minus sign), there can be ambiguity in how we identify traces. 
The simplest example is the fully contracted adjoint quadratic Casimir $f^{a b c} f^{a b c}$. 
In such cases, we must assert a particular index as a loop index, and this in turn will fix two other indices as tracing indices. 
Once we fix indices as loop indices, we will tag them with a ``$*$" to keep track of them. 
Once an index is tagged as a loop index in one $f$, its accompanying contraction must be tagged as a loop index if appearing in another $f$. 
Consequently, the remaining two indices (and their accompanying contractions) in the structure constant must be tracing indices.
We discern tracing indices by underlining them. 
Continuing with our example, if we assert $b$ is a loop index (thus fixing $a$ and $c$ and tracing indices), we find
\begin{align}
f^{a b c} f^{a b c} &= f^{\underline{a} b^* \underline{c}} f^{\underline{a} b^* \underline{c}}\\
&=-f^{b^*  \underline{c}  \underline{a}} f^{b^*  \underline{a}  \underline{c}} \nonumber\\ 
&= -F^{b b}.\nonumber
\end{align} 

In this section, we detail an algorithm on how to systematically identify which indices should be loop or tracing indices. The overall goal is to identify a loop index for each structure constant, but with minimally identifying tracing indices; this is because over-identifying tracing indices will not allow us to properly identify traces. However, this is sometimes not possible and we will have to consider smaller (products of) traces in the term, discarding some $f$'s while trying to trace over as many as possible. The algorithm is as follows.

\begin{enumerate}
\item
Identify $f$'s that only share one contraction with other $f$'s; no longer consider these terms for the trace as $f$'s must be at least doubly contracted to be a part of the trace. Label the one contracted index as a loop index as a result. 
 For example, consider the term $d^{x y z} f^{a y z} f^{a b c} f^{x b c}$; observe that $f^{a y z}$ is only contracted once with the other $f$'s via $a$; thus we do not consider it to be a part of the trace (now coloring it red) and label $a$ as a loop index, as well as $b$ and $c$ as tracing indices.
 \begin{align}
 d^{x y z} f^{a y z} f^{a b c} f^{x b c} & = d^{x y z}  \textcolor{red}{f^{a y z}} f^{a^* \underline{b} \underline{c}} f^{x b c} 
 \end{align}
\item
\label{item:colorstep2}
Next, identify $f$'s that share exactly two contractions with other $f$'s. The contracted indices must be tracing indices and thus we must label the the non-contracted index as a loop index. 
 For the above example, this applies to $f^{x b c}$: we label $b$ and $c$ as tracing indices and $x$ as a loop index and can consequently identify the trace.
 \begin{align}
 d^{x y z} f^{a y z} f^{a b c} f^{x b c} & = d^{x y z}  \textcolor{red}{f^{a y z}} f^{a^* \underline{b} \underline{c}} f^{x^* \underline{b} \underline{c}} \nonumber \\
 & = -d^{x y z}  \textcolor{red}{f^{a y z}} f^{a^* \underline{b} \underline{c}} f^{x^* \underline{c} \underline{b}} \nonumber \\
 & = -d^{x y z}  \textcolor{red}{f^{a y z}} F^{a x} 
 \end{align}
\item
Next, we address $f$'s that share all three contractions with other $f$'s.
This is where the difficulty lies in choosing loop versus tracing indices.
\begin{enumerate}
\item
\label{item:colorstep3a}
First among triply contracted $f$'s, if two contractions are already labeled as tracing indices elsewhere, then they must be tracing indices in the triple contraction term; the leftover index is then labeled as a loop index. 
 For example, consider the triply contracted structure constant $f^{b i_2 i_3}$ in the term
 \begin{align*}
 \dots f^{a^* \underline{i_1} \underline{i_2}} f^{b i_2 i_3} f^{c^* \underline{i_3} \underline{i_4}} f^{b i_5 i_6} \dots,
 \end{align*}
 we label $i_2$ and $i_3$ as tracing indices and $b$ as a loop index: $f^{b^* \underline{i_2} \underline{i_3}}$.

\item
\label{item:colorstep3b}
Next, among triply contracted $f$'s, if only one contraction is already labeled as a tracing index elsewhere, then it must be tracing index in the triple contraction term. Then, if one of the contractions of the remaining two indices also lies in an $f$ with only one tracing index, choose it as a loop index. The leftover third index must be a tracing index. 
 For example, consider the triply contracted structure constants $f^{b i_2 i_3}$ and $f^{b i_4 i_1}$ in the term
  \begin{align*}
 \dots f^{a i_1 i_2} f^{b i_2 i_3} f^{c^* \underline{i_3} \underline{i_4}} f^{b i_4 i_1} \dots.
 \end{align*}
 First, we label $i_3$ and $i_4$ as tracing indices:
 \begin{align*}
 \dots f^{a i_1 i_2} f^{b i_2 \underline{i_3}} f^{c^* \underline{i_3} \underline{i_4}} f^{b \underline{i_4} i_1} \dots.
 \end{align*}
 Then we can identify $b$ as an index of a structure constant with only one labeled tracing index in both triply contracted terms, $f^{b i_2 \underline{i_3}}$ and $f^{b \underline{i_4} i_1}$. We label it as a loop index, and also label $i_2$ and $i_1$ as tracing indices:
  \begin{align*}
 \dots f^{a i_1 i_2} f^{b^* \underline{i_2} \underline{i_3}} f^{c^* \underline{i_3} \underline{i_4}} f^{b^* \underline{i_4} \underline{i_1}} \dots.
 \end{align*}

\item
\label{item:colorstep3c}
If we cannot find a common contraction between triply contracted $f$'s with one tracing index, we simply look for contractions between triply contracted $f$'s with no labeled indices, and assert one as a tracing index. 
We continue to do so until another we can cite another step.

Alternatively, in a completely unlabeled expression with only triply contracted $f$'s, we start by asserting any index as a loop index.

\item
\label{item:colorstep3d}
Finally, if we are unable to discern a loop index and two tracing indices in a triply contracted structure constant, we must discard a doubly contracted $f$ as a candidate for the trace, and look for smaller adjoint traces by beginning the algorithm again. A worked example where this occurs is shown in Section~\ref{sec:ColorExample2}.








\end{enumerate}
\end{enumerate}
\subsubsection{Reduction of Traces in the Adjoint Representation}
\label{sec:ReduceAdjointTrace}
We iterate the above algorithm until all structure constants have one loop index and two tracing indices, or are discarded from the trace. 
After identifying the traces $F$ in an expression, we eliminate contracted indices within the same trace, as done in the fundamental representation. 
The commutation relation in the adjoint representation yields
\beq
\label{eq:Fcommute}
F^{\dots a b \dots} = F^{\dots b a \dots} - f^{a b c} F^{\dots c \dots}.
\eeq
After commuting contracted indices such that they are neighboring in the loop, we use:
\beq
\label{eq:F^aa}
F^{a_1 \dots b b \dots a_n} = -C_A F^{a_1 \dots a_n}.
\eeq
Now, all indices of a single $F$ should be unique.

To reduce traces in the adjoint representation, we separately consider traces with either an odd or even number of indices. 
Due to the antisymmetry of the structure constants and the cyclic property of traces, loops with an odd number of indices $n$ follow:
\beq
\label{eq:oddLoops}
F^{a_1 a_2 \dots a_{n-1} a_n} = -F^{a_1 a_n a_{n-1} \dots a_2}.
\eeq
Then, we can simply commute indices on the right hand side until we can solve for $F^{a_1 a_2 \dots a_{n-1} a_n}$ explicitly in terms of commutator terms with a fewer number of indices.
For example:
\begin{align}
    F^{a_ 1 a_ 2 a_ 3 a_ 4 a_ 5}
    &=-F^{a_ 1 a_ 5 a_ 4 a_ 3 a_ 2}\\
    &=-F^{a_ 1 a_ 2 a_ 3 a_ 4 a_ 5}+f^{a_ 1 a_ 5 k} F^{k a_ 4 a_ 3 a_ 2}+f^{a_ 3 a_ 2 k}F^{a_ 5 a_ 1 a_ 4 k} \nonumber \\
    &~\quad+f^{a_ 4 a_ 2 k}F^{a_ 5 a_ 1 k a_ 3}+f^{a_ 4 a_ 3 k}F^{a_ 5 a_ 1  a_ 2 k}, \nonumber\\
    \Rightarrow 
    F^{a_ 1 a_ 2 a_ 3 a_ 4 a_ 5}&=\frac{1}{2} \left(f^{a_ 1 a_ 5 k} F^{k a_ 4 a_ 3 a_ 2}+f^{a_ 3 a_ 2 k}F^{a_ 5 a_ 1 a_ 4 k}+f^{a_ 4 a_ 2 k}F^{a_ 5 a_ 1 k a_ 3}+f^{a_ 4 a_ 3 k}F^{a_ 5 a_ 1  a_ 2 k}\right).\nonumber
\end{align}
Next, for even number of indices n, we note a similar property:
\beq
F^{a_1 a_2 \dots a_{n-1} a_n} = F^{a_1 a_n a_{n-1} \dots a_2}.
\eeq
We can use the same algorithm that we used in the reduction of fundamental traces, but also combine permutations within the symmetrized trace that are reversals of other permutations, effectively reducing the number of terms we have consider by a factor of two. Additionally, note the totally symmetric tensor in the adjoint representation is given by:
\beq
i^n d_A^{a_ 1 \dots a_n}=\frac{1}{n!}\sum_\pi F^{a_{\pi(1)} \dots a_{\pi(n)}}.
\eeq
The reduction relation for adjoint indices with an even number of indices is:
\beq
\label{eq:evenFReduction}
F^{a_1 \dots a_n} = F^{a_1 \dots a_n}- \frac{1}{n!} \sum_\pi F^{a_{\pi(1)} \dots a_{\pi(n)}} + i^n d_A^{a_1 \dots a_n}. 
\eeq
We end the recursion at a two-index loop, which is related to the adjoint quadratic Casimir: 
\beq
\label{eq:Fab}
F^{a b}=f^{a i_ 1 i_ 2}f^{b i_ 2 i_ 1}=-C_A \delta^{a b}.
\eeq
We also tabulate the three-index loop for speed:
\beq
\label{eq:Fabc2fabc}
F^{a b c}=\frac{1}{2} C_A f^{a b c}.
\eeq
\subsubsection{Additional Properties and Identities}
\label{sec:coloridentities}
As stated before, structure constants can be contracted with symmetric tensors in such a way that they do not form loops. Here, we note identities concerning these cases.
Since structure constants are antisymmetric tensors, double (and triple) contractions with symmetric color tensors vanish:
\beq
\label{eq:fxdvanish}
f^{a b c} d_R^{a_1 \dots a_n b c} = 0.
\eeq
Additionally, following eq.~\ref{eq:oddLoops} symmetric tensors with an odd number of indices $n$ in the adjoint representation also vanish.
\beq
d_A^{a_1 \dots a_n}=0, \hspace{1cm} n\text{ odd}.
\eeq
Finally, there exist nontrivial "triangle" identities, of which we note three classes:
\bea
\label{eq:triangle1}
d_R^{a b i}f^{a j c}f^{b k c} &=& \frac{C_A}{2}d_R^{i j k},\\
d_{R_1}^{i_ 1 j_ 1 \dots j_n k_ 1}d_{R_2}^{i_ 2 j_ 1 \dots j_n k_ 2}f^{k_ 1 
\label{eq:triangle2}
k_ 2 i_ 3} &=& \frac{1}{D_A(n+1)} C^{R_1 R_2}_{n+2} f^{i_ 1 i_ 2 i_ 3}, \\
\label{eq:triangle3}
d_{R_1}^{i_ 1 j_ 1 \dots j_n k_ 1}d_{R_2}^{i_ 3 \dots i_m j_ 1 \dots j_n k_ 
2}f^{k_ 1 k_ 2 i_ 2} &=&-\frac{1}{n+1} d_{R_1}^{j_ 1 \dots j_{n+1} k} 
d_{R_2}^{i_ 3 \dots i_m j_ 1 \dots j_{n+1}} f^{k i_ 1 i_ 2}.
\eea
\subsection{Worked Examples}
In this section, we illustrate two examples to work through the steps of the algorithm.



\subsubsection{Example with Cubic Casimir Invariant}
\begin{figure}[H]
    \centering
            


    \begin{tikzpicture}
        \begin{feynman}[large]
        \vertex (P);
        \vertex[below=of P] (a);
        \vertex[left=of P] (del12);
        \vertex[left=of a] (Ta23);
    
        \vertex[left=of del12] (Tdjk);
        \vertex[left=of Ta23] (Teki);
        \vertex[left=of Tdjk] (fd12);
        \vertex[left=of Teki] (fe34);
        \vertex[left=of fd12] (f256);
        \vertex[left=of fe34] (f478);
        \vertex[left=of f256] (k);
        \vertex[left=of f478] (i);
        \diagram*{
            (P) --[photon] (del12),
            (a) --[gluon, edge label'=\(a_1\)] (Ta23),
            
            (del12) --[fermion, edge label'=\(i_1\)] (Tdjk),
            (Ta23) --[fermion, edge label'=\(i_2\)] (del12),
            (Teki) --[fermion, edge label'=\(i_3\)] (Ta23),
            (Tdjk) --[fermion, edge label'=\(i_4\)] (Teki),

            (Tdjk) --[gluon, edge label'=\(a_3\)] (fd12),
            (Teki) --[gluon, edge label'=\(a_2\)] (fe34),
            (fe34) --[gluon, edge label'=\(a_4\)] (fd12),
            (fd12) --[gluon, edge label'=\(a_5\)] (f256),
            (fe34) --[gluon, edge label'=\(a_6\)] (f478),
            (f478) --[fermion, edge label'=\(j\)] (f256),
            (i) --[fermion, edge label'=\(i\)] (f478),
            (f256) --[fermion, edge label'=\(k\)] (k)
            };
        \end{feynman}
    \end{tikzpicture}
    
    \caption{First worked example, $q\bar{q} \rightarrow \gamma g$ at three loops. All fermion lines are quarks, with fundamental color indices labeled $i,\,j,\,k,\,i_1,\,i_2,\,i_3,$ and $i_4$. Gluons have adjoint color indices labeled $a_1,\,a_2,\,a_3,\,a_4,\,a_5,$ and $a_6$.}
    \label{fig:colordiag1}  
\end{figure}
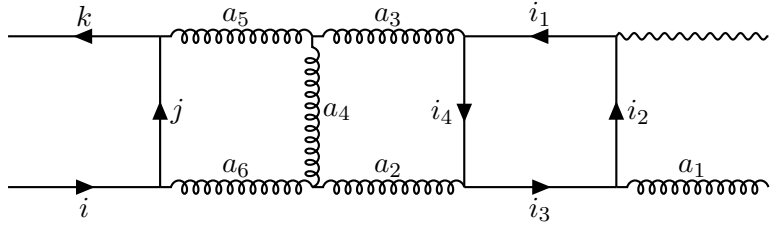
Here, we focus on the diagram in fig.~\ref{fig:colordiag1}, which contributes to Drell-Yan production beginning at N$^4$LO;  we have labeled all the color indices (fundamental and adjoint). Since this diagram has two external legs with fundamental indices $i$ and $k$ and one external leg with adjoint index $a_1$, we project out to a scalar color amplitude by contracting with $t^{a_1}_{i k}$. Then, from QCD Feynman rules and after projection, the color factor for this diagram is:
\beq
\label{eq:D1}
\begin{aligned}
D_1 &= \delta_{i_1 i_2} t^{a_1}_{i_2 i_3} t^{a_2}_{i_3 i_4}  t^{a_3}_{i_4 i_1} t^{a_1}_{i k}t^{a_5}_{k j}t^{a_6}_{j i} f^{a_3 a_4 a_5} f^{a_2 a_6 a_4}\\
&=T^{a_1 a_2 a_3}T^{a_1 a_5 a_6}f^{a_3 a_4 a_5} f^{a_2 a_6 a_4}
\end{aligned}
\eeq
Then, we use eq.~\ref{eq:TrTabc} to reduce our fundamental traces:
\begin{align*}
    D_1 &= \left(d_F^{a_1 a_2 a_3} + \frac{i}{2} T_F f^{a_1 a_2 a_3} \right)\left(d_F^{a_1 a_5 a_6} + \frac{i}{2} T_F f^{a_1 a_5 a_6} \right)f^{a_3 a_4 a_5} f^{a_2 a_6 a_4}.
\end{align*}
We have three terms we can compute separately:
\bea
D_{1,1} &=& d_F^{a_1 a_2 a_3}d_F^{a_1 a_5 a_6}f^{a_3 a_4 a_5} f^{a_2 a_6 a_4},\\
D_{1,2} &=& \frac{i}{2} T_F \left(d_F^{a_1 a_5 a_6}f^{a_1 a_2 a_3}f^{a_3 a_4 a_5} f^{a_2 a_6 a_4}+ d_F^{a_1 a_2 a_3} f^{a_1 a_5 a_6}f^{a_3 a_4 a_5} f^{a_2 a_6 a_4}\right),\\
D_{1,3} &=& -\frac{1}{4}T_F^2 f^{a_1 a_2 a_3}f^{a_1 a_5 a_6}f^{a_3 a_4 a_5} f^{a_2 a_6 a_4}.
\eea
First, we use eq.~\ref{eq:triangle1} on $D_{1,1}$:
\begin{align*}
    D_{1,1} &=-\frac{1}{2}C_A \left(d_F^{a_1 a_2 a_3}\right)^2 \\
    &=-\frac{1}{2}C_A C_3^{FF}.
\end{align*}

Next, for $D_{1,2}$ we refer to step~\ref{item:colorstep2} to identify $a_1$, $a_5$, and $a_6$ as loop indices and $a_2$, $a_4$, and $a_6$ as tracing indices for the first term in the paranthesis; then for the second term, we identify $a_1$, $a_2$, and $a_3$ as loop indices and $a_4$, $a_5$, and $a_6$ as tracing indices. Then, we have:
\begin{align*}
    D_{1,2} &=\frac{i}{2} T_F\left(d_F^{a_1 a_5 a_6}F^{a_1 a_5 a_6}+d_F^{a_1 a_2 a_3}F^{a_1 a_2 a_3}\right) \\
    & =\frac{i}{4} T_F C_A\left(d_F^{a_1 a_5 a_6}f^{a_1 a_5 a_6}+d_F^{a_1 a_2 a_3}f^{a_1 a_2 a_3}\right) \\
    &=0,
\end{align*}
where we used eq~\ref{eq:Fabc2fabc} to reduce the adjoint traces; then, the two terms individually vanish by eq.~\ref{eq:fxdvanish}.

Finally, for $D_{2,3}$ we are free to choose $a_1$ as a loop index by step~\ref{item:colorstep3c}; this consequently labels $a_2$, $a_3$, $a_5$, and $a_6$ as tracing indices by step~\ref{item:colorstep3a}. Then, we must label $a_4$ as a loop index.
\begin{align*}
   D_{1,3} &= -\frac{1}{4}T_F^2 f^{a_1^* \underline{a_2} \underline{a_3}}f^{a_1^* \underline{a_5} \underline{a_6}}f^{a_4^* \underline{a_5} \underline{a_3}}f^{a_4^* \underline{a_2} \underline{a_6}} \\
   & = -\frac{1}{4}T_F^2 F^{a_1 a_4 a_1 a_4} \\
   &= -\frac{1}{4}T_F^2 \left(F^{a_4 a_1 a_1 a_4} -f^{a_1 a_4 b}F^{b a_1 a_4}\right),
\end{align*}
where we used eq.~\ref{eq:Fcommute} to commute $a_1$ and $a_4$. Then, we use eq.~\ref{eq:F^aa}:
\begin{align*}
    D_{1,3} &=-\frac{1}{4}T_F^2 \left(-C_A F^{a_4 a_4} -f^{a_1 a_4 b}F^{b a_1 a_4}\right).
\end{align*}
Then, we use eqs.~\ref{eq:Fab} and~\ref{eq:Fabc2fabc} to reduce the adjoint traces:
\begin{align*}
    D_{1,3} &=-\frac{1}{4}T_F^2\left(C_A^2 \delta^{a_4 a_4} -\frac{1}{2}C_A f^{a_1 a_4 b}f^{b a_1 a_4}\right) \\
    &= -\frac{1}{8}T_F^2 D_A C_A^2,
\end{align*}
by eq.~\ref{eq:adjointQuadraticCasimir}.
This gives the complete result:
\begin{align*}
    D_1 &= D_{1,1}+D_{1,2}+D_{1,3} \\
    & = -\frac{1}{2}C_A C_3^{FF}-\frac{1}{8}T_F^2 D_A C_A^2.
\end{align*}

\subsubsection{Example with Nontrivial Adjoint Structure}
\label{sec:ColorExample2}  
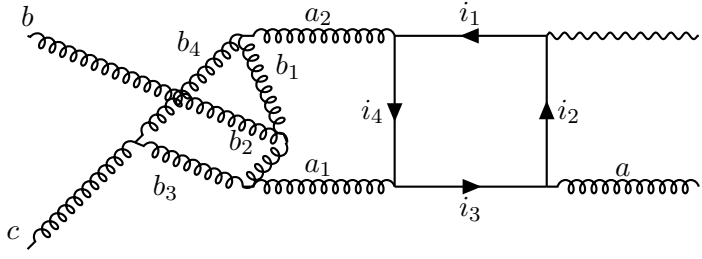
\begin{figure}[H]
    \centering

        



\begin{tikzpicture}
    \begin{feynman}[large]
        \vertex (P);
        \vertex[below=of P] (a);
        \vertex[left=of P] (del12);
        \vertex[left=of a] (Ta23);
    
        \vertex[left=of del12] (Tdjk);
        \vertex[left=of Ta23] (Teki);
        
        \vertex[left=of Tdjk] (fd12);
        \vertex[left=of Teki] (fe34);
        \vertex[left=0.4cm of fd12, shift={(0cm, -0.1cm)}] (label_node) {\(b_4\)};
        \vertex[below left of=fd12] (f56);
        \vertex[right=of f56] (fb35);
        \vertex[above left of=f56, label=\(b\)] (b);
        \vertex[below left of=f56, label=\(c \quad\)] (c);
        
        \diagram*[]{
            (P) --[photon] (del12),
            (a) --[gluon, edge label'=\(a\)] (Ta23),
            
            (del12) --[fermion, edge label'=\(i_1\)] (Tdjk),
            (Ta23) --[fermion, edge label'=\(i_2\)] (del12),
            (Teki) --[fermion, edge label'=\(i_3\)] (Ta23),
            (Tdjk) --[fermion, edge label'=\(i_4\)] (Teki),

            (Tdjk) --[gluon, edge label'=\(a_2\)] (fd12),
            (Teki) --[gluon, edge label'=\(a_1\)] (fe34),
            (fd12) --[gluon, edge label=\(b_1\)] (fb35),
            (fe34) --[gluon, edge label=\(b_2\)] (fb35),
            (fd12) --[gluon] (f56),
            (fe34) --[gluon, edge label=\(b_3\)] (f56),
            (f56) --[gluon] (c),
            (fb35) --[gluon] (b),
        };
    \end{feynman}
\end{tikzpicture}

    \caption{Second worked example, $g g \rightarrow \gamma g$ at three loops. All fermion lines are quarks, with fundamental color indices labeled $i_1,\,i_2,\,i_3,$ and $i_4$. External gluons have adjoint color indices labeled $a,\,b,$ and $c$, while internal gluons have color indices labeled $a_1,\,a_2,\,b_1,\,b_2,\,b_3,$ and $b_4$}
    \label{fig:colordiag2}  
\end{figure}
Now, we focus on the diagram in fig.~\ref{fig:colordiag2}; we have again labeled all the color indices (fundamental and adjoint). This diagram has three external legs with with adjoint indices $a$, $b$, $c$, so we project out to a scalar color amplitude by contracting with $f^{a b c}$. Then, from QCD Feynman rules and after projection, the color factor for this diagram is:
\beq
\begin{aligned}
     D_2 & = \delta_{i_1 i_2} t^{a}_{i_2 i_3} t^{a_1}_{i_3 i_4} t^{a_2}_{i_4 i_1} f^{a_1 b_3 b_2} f^{b b_1 b_2} f^{a_2 b_1 b_4} f^{c b_4 b_3} f^{a b c} \\ &= T^{a a_1 a_2} f^{a_1 b_3 b_2} f^{b b_1 b_2} f^{a_2 b_1 b_4} f^{c b_4 b_3} f^{a b c}   
\end{aligned}
\eeq
Then, we use eq.~\ref{eq:TrTabc} to reduce the fundamental trace:
\begin{align*}
    D_2 = & \left(d_F^{a a_1 a_2} + \frac{i}{2}T_F f^{a  a_1 a_2}\right)  \\ & \times \left(f^{a_1 b_3 b_2} f^{b b_1 b_2} f^{a_2 b_1 b_4} f^{c b_4 b_3} f^{a b c}\right).
\end{align*}
We deal with these two terms separately, identifying adjoint traces. First:
\beq
D_{2,1} = d_F^{a a_1 a_2}f^{a_1 b_3 b_2} f^{b b_1 b_2} f^{a_2 b_1 b_4} f^{c b_4 b_3} f^{a b c}.
\eeq
First, we label $a$, $a_1$, and $a_2$ as loop indices according to step~\ref{item:colorstep2}:
\begin{align*}
    D_{2,1} = d_F^{a a_1 a_2}f^{a_1^* \underline{b_3} \underline{b_2}} f^{b b_1 b_2} f^{a_2^* \underline{b_1} \underline{b_4}} f^{c b_4 b_3} f^{a^* \underline{b} \underline{c}}.
\end{align*}

Then, we can see that by step~\ref{item:colorstep3a}, all the indices of $f^{b b_1 b_2}$ and $f^{c b_4 b_3}$ must be tracing indices; but this is not allowed, there must only be two tracing indices and one loop index in each $f$. Then, according to step~\ref{item:colorstep3d}, we discard the doubly contracted $f^{a b c}$, and can now identify $b_1$, $b_2$, $b_3$, and $b_4$ as tracing indices by step~\ref{item:colorstep3a}, as well as $b$ and $c$ as loop indices. So we have:
\begin{align*}
    D_{2,1} &= d_F^{a a_1 a_2}f^{a_1^* \underline{b_3} \underline{b_2}} f^{b^* \underline{b_1} \underline{b_2}} f^{a_2^* \underline{b_1} \underline{b_4}} f^{c^* \underline{b_4} \underline{b_3}}  \textcolor{red}{f^{a b c}}
    \\ &=-d_F^{a a_2 a_3} F^{a_1 b a_2 c}  \textcolor{red}{f^{a b c}}.
\end{align*}
Next, we reduce the four-index adjoint trace with eq.~\ref{eq:evenFReduction}:
\begin{align*}
    D_{2,1} &=-d_F^{a a_1 a_2}\left(\frac{1}{6} C_A f^{a_1 d c} f^{a_2 b d}+\frac{1}{6} C_A f^{a_1 b d} f^{c a_2 d}+d_A^{a_1 a_2 b c}\right)f^{a b c}.
\end{align*}
We have $d_A^{a_1 a_2 b c}f^{a b c}=0$ by eq.~\ref{eq:fxdvanish}. Now, we can label $a$, $a_1$, and $a_2$ as loop indices with no issues, and identify the following traces:
\begin{align*}
    D_{2,1} &= -\frac{1}{6}C_A d_F^{a a_1 a_2}\left(F^{a a_1 a_2}+F^{a a_2a_1}\right) \\
    & =-\frac{1}{12}C_A^2 d_F^{a a_1 a_2}\left(f^{a a_1 a_2}+f^{a a_2 a_1}\right) \\
    &= 0,
\end{align*}
again by eq.~\ref{eq:fxdvanish} (and also because the term in the parenthesis vanishes by antisymmetry).

Next, we address:
\beq
D_{2,2} =\frac{i}{2}T_F f^{a a_1 a_2} f^{a_1 b_3 b_2} f^{b b_1 b_2} f^{a_2 b_1 b_4} f^{c b_4 b_3} f^{a b c}.
\eeq

We are free to choose $a$ as a loop index by step~\ref{item:colorstep3c}; this consequently labels $a_1$, $a_2$, $b$, and $c$ as tracing indices by step~\ref{item:colorstep3a}.
\begin{align*}
    D_{2,2} =\frac{i}{2}T_F f^{a^* \underline{a_1} \underline{a_2}}f^{a^* \underline{b} \underline{c}} f^{\underline{a_1} b_3 b_2} f^{\underline{b} b_1 b_2} f^{\underline{a_2} b_1 b_4} f^{\underline{c} b_4 b_3}.
\end{align*}
Then, since the index $b_1$ appears twice in structure constants with only one tracing index labeled ($b$ in the first and $a_2$ in the second), we can label it as a tracing index by step~\ref{item:colorstep3b}. This implies that $b_2$ and $b_4$ must be loop indices, which in turn determines that $b_3$ is a tracing index. As a result, we have:
\begin{align*}
    D_{2,2} &=\frac{i}{2}T_F f^{a^* \underline{a_1} \underline{a_2}}f^{a^* \underline{b} \underline{c}} f^{b_2^* \underline{a_1} \underline{b_3}} f^{b_2^* \underline{b} \underline{b_1}} f^{b_4^* \underline{a_2} \underline{b_1}} f^{b_4^* \underline{b_3} \underline{c}} \\
    &=-\frac{i}{2}T_F F^{a b_4 b_2 a b_4 b_2}.
\end{align*}
After commuting like indices such that they are adjacent using eq.~\ref{eq:Fcommute}, and then contracting adjacent indices according to eq.~\ref{eq:F^aa}, we can continue to iterate the algorithm until we finally arrive at:

\begin{align*}
    D_{2,2}=0,
\end{align*}
which ultimately gives:
\begin{align*}
    D_2&=D_{2,1}+D_{2,2} \\
    &=0.
\end{align*}

\subsubsection{Example Usage in \textsc{Mathematica}}

We provide \textsc{Mathematica} documentation for \textsc{Cifar} in Tables~\ref{tab:cifarObjects} and~\ref{tab:cifarFunctions}. A code example evaluating the color factor $D_1$ (eq.~\ref{eq:D1}) from the previous section is shown below.

\exbox{
\inline{1}{&{\tt D1 = deltaF[i1,i2]*}\\
&{\tt TT[\{a1\},i2,i3]*TT[\{a2\},i3,i4]*TT[\{a3\},i4,i1]*}\\
&{\tt TT[\{a1\},i,k]*TT[\{a5\},k,j]*TT[\{a6\},j,i]*}\\
&{\tt ff[a3,a4,a5]*ff[a2,a6,a4];}}\\
\inline{2}{&{\tt D1CIFAR = CIFARReduce[ D1 ]}}\\
\breakline \\
\outline{2}{&{\tt -1/2*(C3FF*CA) - (CA\^{}2*DA*TF\^{}2)/8}}
}
Then, we can check if the results are consistent when evaluating explicitly in $\text{SU}(n_c)$.
\exbox{
\inline{3}{&{\tt SUncReduce[ D1 ]}}\\
\inline{4}{&{\tt SUncReduce[ D1CIFAR ] // Expand}}\\
\breakline \\
\outline{3}{&{\tt -1/8 + (3*nc\^{}2)/16 - nc\^{}4/16}}\\
\outline{4}{&{\tt -1/8 + (3*nc\^{}2)/16 - nc\^{}4/16}}
}

\begin{table}[H]
\bgfb
\multicolumn{2}{c}{\textbf{Table~\ref{tab:cifarObjects}: \textsc{Cifar} Objects}}\\
\multicolumn{2}{l}{(Products and traces of products of) members of the color group:}
\\
\\
\texttt{ff[a,b,c]} & Structure constants $f^{a b c}$ as defined in eq.~\ref{eq:lieCommutation}.
\\
\\
\texttt{TT[\{a\},i,j]} & Color group generators in the fundamental representation, $t^a_{i j}$; note the adjoint index \texttt{a} in the brackets.
\\
\\
\texttt{TT[\{a1, \dots, an\},i,j]} & Products of fundamental generators $T^{a_1 \dots a_n}_{i j}$ as defined in \ref{eq:fundProduct}. To express traces of these products $T^{a_1 \dots a_n}$ as defined in eq.~\ref{eq:fundLoop}, one may simply input identical arguments for the entry indices: \texttt{T[\{a1, \dots, an\},i,i]}.
\\
\\
\texttt{FF[\{a1, \dots, an\}]} & Traces in the adjoint representation $F^{a_1 \dots a_n}$, or "loops", as defined in eq~\ref{eq:Floop}.
\\
\\
\multicolumn{2}{l}{Other associated tensors:}
\\
\\
\texttt{dF[a1, \dots, an]} \linebreak \texttt{dA[a1, \dots, an]}  & Symmetric color tensors $d_R^{a_1 \dots a_n}$ in the fundamental and adjoint representations respectively as defined by eq.~\ref{eq:symcolortensors}.
\\
\\
\texttt{deltaF[i,j]} \linebreak \texttt{deltaA[a,b]}  & Kronecker delta symbols in the fundamental ($\delta_{i j}$) and adjoint ($\delta^{a b}$) representations respectively. 
\\
\\
\multicolumn{2}{l}{Constants of the color group:}
\\
\\
\texttt{DF} \linebreak \texttt{DA} & Dimension of the fundamental ($D_F$) and adjoint ($D_A$) representation respectively.
\\
\\
\texttt{TF} & Quadratic Dynkin index of the fundamental representation $T_F$, as defined in eq.~\ref{eq:quadraticDynkinIndex}.
\\
\\
\texttt{CF} \linebreak \texttt{CA} & Quadratic Casimir invariant of the fundamental ($C_F$) and adjoint ($C_A$) representation respectively, as defined in eqs.~\ref{eq:quadraticCasimir} and~\ref{eq:adjointQuadraticCasimir}.
\\
\\
\texttt{C3FF} \linebreak 
\texttt{C4FF} \linebreak 
\texttt{C4AF} \linebreak 
\texttt{C4AA} \linebreak 
& Higher order cubic ($C_3^{FF}$) and quartic ($C_4^{FF}$, $C_4^{AF}$, $C_4^{AA}$) Casimir invariants as defined in eq~\ref{eq:generalCasimir}.
\\
\\
\texttt{nc} \linebreak 
& The number of colors $n_c$ corresponding specifically to the color group $\text{SU}(n_c)$.
\\
\egfb
\captionsetup{labelformat=empty} 
\caption{\label{tab:cifarObjects}}
\end{table}

\begin{table}[H]
\bgfb
\multicolumn{2}{c}{\textbf{Table~\ref{tab:cifarFunctions}: \textsc{Cifar} Functions}}\\
\multicolumn{2}{l}{Primary functions:}
\\
\texttt{CIFARReduce[expr]} & Reduces fully contracted expressions in the color group in \texttt{expr} as outlined throughout this section; the result is in terms of Casimir invariants and dimensions of the fundamental and adjoint representations.
\\
\\
\texttt{SUncReduce[expr]} & Reduces fully contracted expressions in the color group in \texttt{expr}, assuming $\text{SU}(n_c)$; the result is in terms of the number of colors $n_c$. Also expresses Casimir invariants and dimensions of various representations in terms $n_c$ as according to eqs.~\ref{eq:SUncCasimir} and~\ref{eq:SUncDims2}.
\\
\\
\multicolumn{2}{l}{Auxiliary functions:}
\\
\texttt{StrF[\{a1, ..., an\}]} \linebreak \texttt{StrA[\{a1, ..., an\}]} & Returns the explicit symmetrized trace $\text{Str}[t_R^{a_1} \dots t_R^{a_n}]$ in the fundamental and adjoint (in terms of structure constants) representation respectively.
\\
\\
\texttt{ToTTProductForm[expr]} & Identifies products of fundamental generators in \texttt{expr} and expresses them in terms of product objects \texttt{TT[\{a1, ..., an\}, i, j]}.
\\
\\
\texttt{ContractTT[TT[\{a1, ..., an\},i,j]]} & Reduces contractions of adjoint indices within \texttt{\{a1, ..., an\}}.
\\
\\
\texttt{ContractffTT[expr]} & Reduces contractions of doubly contracted adjoint indices between products of fundamental generators and structure constants in \texttt{expr} as outlined in eq.~\ref{eq:ContractffTT}.
\\
\\
\texttt{TraceTT[TT[\{a1, ..., an\},i,i]]]} & Reduces the trace of products of fundamental generators $T^{a_1 \dots a_n}$ in terms of structure constants and symmetric color tensors of lower degree as outlined in the remainder of Section~\ref{sec:ReduceTT}.
\\
\\
\texttt{AdjointReduce[expr]} & Recursively reduces fully contracted expressions in the color group with only adjoint objects (including symmetric color tensors $d$ in either representation) in \texttt{expr}. Comprised of the following auxiliary functions:
\\
\\
\texttt{ToFFLoopForm[expr]} & Identifies traces of structure constants in \texttt{expr} and expresses them in terms of loop objects \texttt{FF[{a1, ..., an}]}. Automatically evaluates adjoint quadratic Casimir invariants (eq.~\ref{eq:adjointQuadraticCasimir}).
\\
\\
\texttt{ContractFF[FF[\{a1, ..., an\}]]} & Reduces contractions of adjoint indices within \texttt{\{a1, ..., an\}}.
\\
\\
\texttt{TraceFF[FF[\{a1, ..., an\}]]} & Reduces the adjoint loop $F^{a_1 \dots a_n}$ in terms of structure constants and symmetric color tensors of lower degree.
\\
\\
\egfb
\captionsetup{labelformat=empty} 
\caption{\label{tab:cifarFunctions}}
\end{table}
\section{Conclusions}
\label{sec:conclusions}

We present a key contribution to the inclusive gluon-fusion Higgs-boson and Drell-Yan production cross sections at hadron colliders at N$^4$LO in perturbative QCD.
Specifically, we consider the $\text{RVV}\times \text{V}$ contribution to the partonic cross section - these are the contributions involving a single parton in the final state and one-loop amplitudes interfered with two-loop amplitudes. 
For Higgs production via gluon fusion, we work in an effective theory in which an infinitely massive top quark couples directly to gluons. 
The first of the main results of this article are analytical expressions for the  $\text{RVV}\times \text{V}$ contributions to N$^4$LO corrections to the partonic coefficient functions.
We provide these results as ancillary files along with the arXiv submission of this article.

To achieve our results, we compute scattering amplitudes for a Higgs boson or a virtual photon and three additional partons at one-loop and two-loop level in perturbative QCD. 
Our approach relies on modern scattering amplitude technology and we emphasize in particular our method of computing color factors for our scattering amplitudes.
In particular, we develop a new \textsc{Mathematica} package to compute these factors in terms of invariants of a general compact Lie algebra. 
Next, we compute the required interferences and integrate over phase space directly order by order in the dimensional regulator $\epsilon$. 
The result is a Laurent expansion in $\epsilon$ up to finite order in terms of analytic functions of the invariant mass of the Higgs boson or virtual photon as well as the partonic center-of-mass energy. 

The second main result of this article is our new code package \textsc{Cifar}.
This package implements an algorithm to compute contractions of color generators of an arbitrary Lie algebra.
We present details of the implemented algorithm as well as several examples and use cases of this code.
Our \textsc{Mathematica} implementation of this algorithm is available in the ancillary files attached with the arXiv submission of this article.

In this article, we take a decisive first step towards the computation of the inclusive Higgs-boson and Drell-Yan production cross section at N$^4$LO.
The technology developed for the purposes of this article will be the foundation of this larger goal. 
Our explicit result is a well-defined and essential building block of N$^4$LO cross sections.


\acknowledgments
BM and AS are supported by the United States Department of Energy, Contract DE-AC02-76SF00515. 

\newpage
\appendix
\addcontentsline{toc}{section}{References}
\bibliographystyle{jhep}
\bibliography{paper}

\clearpage
\end{document}